\documentclass{article}

\usepackage{graphicx}
\setkeys{Gin}{width=12cm}   

\begin{document}

\title{Nonextensive Statistics and Multiplicity Distribution in Hadronic Collisions}

\author{C.E. Aguiar and T. Kodama \\ \\
Instituto de F\'{\i}sica, Universidade Federal do Rio de Janeiro \\
Cx.P. 68528, Rio de Janeiro, 21945-970, RJ, Brasil}

\date{}

\maketitle

\begin{abstract}
The multiplicity distribution of particles in relativistic gases is studied in
terms of Tsallis' nonextensive statistics. For an entropic index $q>1$ the
multiplicity distribution is wider than the Poisson distribution with the same
average number of particles, being similar to the negative binomial
distribution commonly used in phenomenological analysis of hadron production
in high-energy collisions.
\end{abstract}

\section{Introduction}

It is always amusing to recognize that the concepts of equilibrium statistical
physics, and hence of thermodynamics, can be applied in (at first sight)
unexpected branches of physics. For example, it has long been known that
several features of multiparticle production in high-energy hadronic
collisions are well described by thermostatistical models
\cite{Fermi,Hagedorn1,Hagedorn2}. More recently, it has been shown that a
simple thermal description of particle abundances works quite well for
the hadrons produced in elementary electron-positron collisions
\cite{Becattini1,Becattini2}. This indicates that the hadronic system
reaches statistical equilibrium, at least for some
observables. One might try to justify this in terms of the large number of
degrees of freedom in the final state.

Thermostatistics is also an important issue in relativistic heavy ion physics.
The main focus of recent investigations in this field is
the production and identification of the quark-gluon plasma, a phase of
nuclear matter consisting of unconfined quarks and gluons. Thermostatistical
aspects play a central role in these investigations, as it is expect that some
form of statistical equilibrium is attained in high-energy collisions of two heavy 
nuclei, at least in the central region \cite{Letessier,BraunM}.

One of the signatures of  ``thermal'' multiparticle production is the
exponential form of the transverse energy distribution of the produced
hadrons. We are tempted to interpret the slope parameter of this distribution
in terms of a temperature of the final state. In fact, up to the ISR energy
region, the transverse energy distribution is consistent with such an
interpretation. The same happens to the observed hadron abundances.

However, at higher energies, the thermal interpretation of the transverse
energy spectra should be modified. The SPS experiments revealed a significant
deviation from the exponential transverse energy distribution, together with
the violation of several scaling laws. Usually, these aspects are understood
in terms of the onset of hard QCD processes such as the formation of
mini-jets. Furthermore, even if local thermal equilibrium is attained,
dynamical effects such as collective flow may entangle with the thermal
transverse spectra. Consequently, for these energies, a pure statistical
description to the overall system cannot be applied.

Recently, it was pointed out \cite{Ignacio,Quarati1,Wilk1,Beck} that the
nonexponential behavior of the transverse energy distribution in high-energy
multiparticle production processes (including electron-positron annihilation)
can be described rather well by the nonextensive thermostatistics of Tsallis
\cite{Tsallis1}. In addition to the excellent quality of the fit to transverse
energy spectra, an interesting point of the extended statistical approach to
multiparticle production is that the ``temperature'' stays almost constant,
independent of the incident energy, conveniently recovering the classical
Hagedorn scenario \cite{Hagedorn1,Hagedorn2,Ignacio}. This opens the
possibility of reviving the statistical description of the complex dynamics of
hadronic collisions in a unified way, including the onset of semi-hard QCD
processes, provided that we generalize the concept of statistical equilibrium.
Differently from the standard statistical mechanics of Boltzmann and Gibbs,
the Tsallis nonextensive thermostatistics contains one extra parameter, the
``entropic index'' $q$, originated in the definition of the entropy. The
physical origin of this parameter is not yet completely understood and a lot
of discussions are in course. It can be attributed to various factors, like
the dynamical suppression of certain domain of the phase space (multifractal
structure near the critical point), the presence of long range forces, or
fluctuations in a small system \cite{Tsallis2,Tsallis3}. For an entropic index
$q=1$ Tsallis statistics reduces to the Boltzmann-Gibbs theory. If $q>1$, rare
events are enhanced relative to the Boltzmann-Gibbs case. If $q<1$, frequent
events are privileged. So far, all phenomenological applications of Tsallis
statistics to high-energy collisions have found $q>1$ in these processes. Some
discussion on the possible origin of nonextensive behavior in hadronic systems
can be found in Refs.~\cite{Wilk2,Rafelski}.

The transverse energy distribution is not the only probe of the formation of a
locally thermalized source of particles in high-energy collisions. Another
important observable is the multiplicity distribution of the produced hadrons.
The negative binomial distribution \cite{Giovannini}
\begin{equation}
P_{N}=\frac{\Gamma(N+k)}{\Gamma(N+1)\Gamma(k)}\,
\frac{(\overline{N}/k)^{N}}{( 1+\overline{N}/k)^{N+k}}\;,
\label{negativeB}
\end{equation}
is often used to express hadron abundances in high-energy processes, where $N$
is the number of particles and $\overline{N}$ the average multiplicity. The
$k$ parameter is related to the variance 
$D^{2}=\overline{N^{2}}-\overline{N}^{2}$ of the distribution by 
\begin{equation}
D^{2}=\overline{N}+\frac{\overline{N}^{2}}{k}\;. 
\label{NB-k}
\end{equation}
Although the negative binomial distribution is known to arise from some
specific processes, such as Bose-Einstein particles with different sources
\cite{Carruthers}, its use in high-energy multiparticle production is rather 
phenomenological.

In this paper, we discuss how the generalized statistical mechanics of Tsallis
(with $q>1$) affects the multiplicity distribution in hadronic collisions. 
We show that for $q$ close to unity the distribution of particle 
numbers in a high-energy gas is approximately negative binomial. 
The Boltzmann-Gibbs and Tsallis statistics are applied to a simple meson 
gas model, and the results are compared to experimental data.

\section{Nonextensive $q$ statistics\label{sec-nonextens}}

First, let us review briefly Tsallis' statistics and derive some formulas
necessary for our discussion. Tsallis generalized the usual Boltzmann-Gibbs
thermostatistics by introducing the ``$q$-entropy''
\begin{equation}
S=\frac{1-\sum_{a}\, p_{a}^{q}}{q-1} \;, 
\label{Sq}
\end{equation}
where $p_{a}$ is the probability associated with microstate $a$, normalized
as
\begin{equation}
\sum_{a}p_{a}=1 \;. 
\label{norm}
\end{equation}
Tsallis also introduced ``$q$-biased'' averages of observables,
\begin{equation}
\langle O \rangle=\frac{1}{\mathcal{C}_{q}}\sum_{a}O_{a}\,p_{a}^{q} \;,
\label{av-O}
\end{equation}
the normalization factor being
\begin{equation}
\mathcal{C}_{q}=\sum_{a}p_{a}^{q}\;.
\end{equation}
The $q$-biased\ microstate probability
\begin{equation}
\widetilde{p}_{a}=\frac{\,p_{a}^{q}}{\mathcal{C}_{q}} 
\label{q-prob} 
\end{equation}
is the probability to be used in the calculation of physical quantities. The
entropic index $q$ is a real number, and in the limit $q\rightarrow1$ the
Boltzmann-Gibbs-Shannon entropy is recovered. In this limit the $q$-average
also reduces to the usual one. The equilibrium probabilities $p_{a}$ are
determined by maximizing the entropy under appropriate constraints. These are
the normalization condition (\ref{norm}), and the fixed (average) value of the
energy $E$ and of any other relevant conserved quantity. We assume for
simplicity that there is only one such quantity, a conserved charge $Q$. The
variational principle is then
\begin{equation}
\delta S+\alpha\sum_{a}\delta p_{a}-\beta_{T}\,\delta E+\gamma\,\delta Q=0 \;,
\label{var}
\end{equation}
where
\begin{equation}
E  =  \sum_{a} E_{a}\widetilde{p}_{a} \;, 
\end{equation}
\begin{equation}
Q  =  \sum_{a} Q_{a}\widetilde{p}_{a}\;,
\end{equation}
and $E_{a}$ and $Q_{a}\;$are the energy and charge of the microstate $a$. The
constants $\alpha$, $\beta_{T}$ and $\gamma$ are Lagrange multipliers. The
later two are associated with the temperature $T$ and chemical potential $\mu$
(in the sense of the second law of thermodynamics) as
\begin{equation}
\beta_{T}=\frac{1}{T}=\left(  \frac{\partial S}{\partial E}\right)_{Q,V}\;,
\end{equation}
\begin{equation}
\gamma=\frac{\mu}{T}\;=-\left(  \frac{\partial S}{\partial Q}\right)_{E,V}\;,
\end{equation}
where $V$ is the volume occupied by the system. Solving the variational
equation (\ref{var}) we obtain the Tsallis distribution,
\begin{equation}
\widetilde{p}_{a}=\frac{1}{Z_{q}}\left\{  \exp_{q}\left[  -\beta\left(
E_{a}-\mu Q_{a}\right)  \right]  \right\}  ^{q}, \label{Pi}%
\end{equation}
where we have defined the ``$q$-exponential function'', $\exp_{q}$, by
\begin{equation}
\exp_{q}(A)  \equiv [ 1 - (q-1)  A ]^{-1/(q-1)} \;. 
\label{Expq}
\end{equation}
We have also introduced the generalized partition function $Z_{q}$ as
\begin{equation}
Z_{q}(\beta,\mu,V) = \sum_{a} \left\{  
  \exp_{q} [ -\beta (  E_{a}-\mu Q_{a} ) ]  \right\}^{q} \;. 
\label{partition}
\end{equation}
Note that in the limit of $q\rightarrow1$ we have
\begin{equation}
\lim_{q\rightarrow1}\exp_{q}\left(  A\right)  = e^{A}
\end{equation}
so that Eqs.~(\ref{Pi}) and (\ref{partition}) reduce to the corresponding
quantities of the usual Boltzmann-Gibbs statistical mechanics. The parameter
$\beta$ appearing in these equations is not the inverse temperature Lagrange 
multiplier $\beta_{T}=1/T$; $\beta$ is related to temperature $T$ by
\begin{equation}
T = \frac{ \beta^{-1} + (q-1)  (E - \mu Q)  }{1+ (1-q)  S}\;. 
\label{T}
\end{equation}
The Tsallis non-intensive temperature $T$ meets some difficulties when 
confronted with the zeroth law of thermodynamics. This point has been 
investigated recently \cite{Abe}, and it was shown that the quantity
\begin{equation}
\widetilde{T}=\beta^{-1}+\left(  q-1\right)  \left(  E-\mu Q\right)  \;,
\label{Tphys}
\end{equation}
sometimes called the ``physical'' temperature, can provide a better 
characterization of thermal equilibrium.

Instead of using the chemical potential $\mu$ to control the average charge
$Q$, as in the grand canonical approach outlined above, it is sometimes
necessary to impose charge conservation exactly. This is particularly
important for small values of $Q$, when fluctuations around the average become
significant. Then a canonical treatment is preferable, in which case the
generalized partition function for a fixed charge $Q$ is given by%
\begin{equation}
Z_{q}(\beta,Q,V)=\sum_{a}\delta\left(  Q-Q_{a}\right)  \left[  \exp_{q}\left(
-\beta E_{a}\right)  \right]  ^{q} \label{canonic-Z}%
\end{equation}
where $\delta(Q-Q_{a})$ is a Kronecker delta. The generalized canonical
probability is%
\begin{equation}
\widetilde{p}_{a}=\frac{\delta(Q-Q_{a})}{Z_{q}(\beta,Q,V)}\left[  \exp
_{q}\left(  -\beta E_{a}\right)  \right]  ^{q}\;. \label{canonic-p}%
\end{equation}

\section{Integral representation of the Tsallis distribution 
\label{sec-integral}}

For further development of the theory, it is useful to introduce the following
integral representation, valid for $q>1$,
\begin{equation}
\left[  \exp_{q}(A)\right]  ^{q}=\int_{0}^{\infty}dxG\left(  x\right)  e^{xA},
\label{int-rep}%
\end{equation}
where
\begin{equation}
G\left(  x\right)  =\frac{\left(  \nu x\right)  ^{\nu}}{\Gamma\left(
\nu\right)  }e^{-\nu x} \label{G(x)}%
\end{equation}
with $\nu=1/\left(  q-1\right)  $. Since $G\left(  x\right)  \geq0$ and
\begin{equation}
\int_{0}^{\infty}G\left(  x\right)  dx=1,
\end{equation}
the function $G\left(  x\right)  $ can be considered as the probability
distribution of the variable $x\in\lbrack0,\infty)$. The maximum of $G(x)$ is
at $x=1$, and the moments of $x$ are
\begin{equation}
\overline{x^{n}}=\int_{0}^{\infty}x^{n}G\left(  x\right)  dx=\frac
{(\nu+n)\left(  \nu+n-1\right)  \cdots\left(  \nu+1\right)  }{\nu^{n}}\;.
\end{equation}
In particular, we have
\begin{equation}
\overline{x} =  \frac{\nu+1}{\nu}=q  \;,
\end{equation}
\begin{equation}
\overline{x^{2}} = \frac{(\nu+1) (\nu+2)}{\nu^{2}}=q(2q-1) \;,
\end{equation}
so that the width of $G(x)$ is
\begin{equation}
\sigma_{x}=\sqrt{\overline{x^{2}}-\overline{x}^{2}}=\sqrt{q (q-1)}\;.
\end{equation}
For $q\rightarrow1$, $G\left(  x\right)  $ tends to the Dirac delta function
$\delta\left(  x-1\right)  $.

Using the integral representation (\ref{int-rep}), we can express the
generalized partition function (\ref{partition}) as
\begin{equation}
Z_{q}(\beta,\mu,V)=\int_{0}^{\infty}dxG(x)Z\left(  x\beta,\mu,V\right)  ,
\label{part-2}%
\end{equation}
where
\begin{equation}
Z\left(  \beta,\mu,V\right)  =\sum_{a}\exp\left[  -\beta\left(  E_{a}-\mu
Q_{a}\right)  \right]  \label{Z0}%
\end{equation}
is the Boltzmann-Gibbs partition function.

The integral representation of the $q$-exponential function is useful for
developing approximations when $q$ is close to unity. As we have seen, in this
case the function $G\left(  x\right)  $ is sharply peaked at $x=1$. Thus, in
any expression of the form
\begin{equation}
I=\int_{0}^{\infty}dxG\left(  x\right)  e^{f\left(  x\right)  }\;,
\end{equation}
we may expand the exponent $f\left(  x\right)  $ around $x=1$ as
\begin{equation}
f\left(  x\right)  \simeq f(1)+f^{\prime}(1)\left(  x-1\right)  +\cdots\;,
\end{equation}
and obtain
\begin{eqnarray}
I  &  \simeq & e^{f(1)-f^{\prime}(1)}\int_{0}^{\infty}dxG\left(  x\right)
e^{xf^{\prime}(1)}\nonumber\\
&  = & e^{f(1)-f^{\prime}(1)}\left\{  \exp_{q}\left[  f^{\prime}(1)\right]
\right\}  ^{q}\nonumber\\
&  \simeq & \exp\left\{  f(1)+(q-1)f^{\prime}(1)\left[  1+f^{\prime}(1)/2\right]
\right\}  \;. \label{q1}%
\end{eqnarray}
In the last step we have used
\begin{equation}
\left[  \exp_{q}(A)\right]  ^{q}\simeq\exp\left[  A+(q-1)\left(
A+A^{2}/2\right)  \right]  \;, \label{expq1}%
\end{equation}
valid for $q\simeq1$. 
Applying these approximations to Eq.~(\ref{part-2}), the generalized 
partition function $Z_{q}(\beta,\mu,V)$ can be written in terms 
of the Boltzmann-Gibbs function $Z(\beta,\mu,V)$ as
\begin{equation}
Z_{q}(\beta,\mu,V)\simeq Z(\beta,\mu,V)\exp\left\{(q-1)\left[ -\beta 
( E-\mu Q ) + \beta^{2} ( E-\mu Q )^{2}/2 \right] \right\}  \;,
\end{equation}
where
\begin{equation}
E-\mu Q=-\frac{\partial\ln Z}{\partial\beta}\;.
\end{equation}

Integral representations of $\exp_{q}$ also exist for $q<1$ \cite{Prato,Lenzi}. 
We will not explore these here because, as mentioned in the Introduction,
$q>1$ seems to be the case of interest in high-energy collisions.

\section{Particle multiplicity distribution}

The probability that the system has exactly $N$ particles is given, in the
Tsallis statistics, by
\begin{equation}
P_{N}=\sum_{a} \delta(N-N_{a})\,\widetilde{p}_{a}%
\end{equation}
where $N_{a}$ is the particle number in state $a$. Defining the partition
function restricted to $N$ particles,
\begin{equation}
Z_{q}^{(N)}(\beta,\mu,V)=\sum_{a}{}\delta(N-N_{a})\left\{  \exp_{q}\left[
-\beta\left(  E_{a}-\mu Q_{a}\right)  \right]  \right\}^{q} \;, 
\label{Zn*}%
\end{equation}
we have
\begin{equation}
P_{N}=\frac{Z_{q}^{(N)}}{Z_{q}}\;. \label{Pn}%
\end{equation}
Using the integral representation of the $q$-exponential function,
we can write
\begin{equation}
Z_{q}^{(N)}(\beta,\mu,V)=\int_{0}^{\infty}dxG\left(  x\right)  Z^{(N)}\left(
x\beta,\mu,V\right)  \label{Zn}%
\end{equation}
where $Z^{(N)}$ is the Boltzmann-Gibbs restricted 
partition function
\begin{equation}
Z^{\left(  N\right)  }\left(  \beta,\mu,V\right)  =\sum_{a}{}\delta
(N-N_{a})\exp\left[  -\beta\left(  E_{a}-\mu Q_{a}\right)  \right]  \;.
\end{equation}

It is useful to introduce the generating function for the multiplicity
distribution, defined by
\begin{equation}
F\left(  t\right)  \equiv\sum_{N=0}^{\infty}t^{N}P_{N}\;.
\end{equation}
For example, if $P_{N}$ is a Poisson distribution,
\begin{equation}
P_{N}=\frac{\overline{N}^{N}}{N!}e^{-\overline{N}}, \label{Poisson}%
\end{equation}
then its generating function is an exponential,
\begin{equation}
F_{P}\left(  t\right)  =\exp[\overline{N}(t-1)]. \label{Gen_pois}%
\end{equation}
For the negative binomial distribution, Eq.~(\ref{negativeB}), we
get
\begin{equation}
F_{NB}\left(  t\right)  =\left[  1-\frac{\overline{N}}{k}\left(  t-1\right)
\right]  ^{-k}=\exp_{q}\left[  \overline{N}\left(  t-1\right)  \right]  \;,
\label{Gen_negb}
\end{equation}
where $q=1+1/k$. It is suggestive to note that the negative binomial
generating function is obtained substituting the exponential in the Poisson
generating function by the $q$-exponential. For large values of $k$ the
negative binomial generating function has the asymptotic behavior%
\begin{equation}
F_{NB}(t)\simeq\exp\left[  \overline{N}(t-1)+\frac{\overline{N}^{2}}%
{2k}(t-1)^{2}+\cdots\right]  \;, \label{NBasymp}%
\end{equation}
a result that will be useful later on. We see that in the limit 
$k\rightarrow \infty$ the negative binomial reduces to the 
Poisson distribution.

\section{Relativistic ideal gas \label{sec-ideal}}

In order to study the Tsallis statistics of a hadronic gas, it is tempting to
apply the results of the previous sections to an ideal relativistic Boltzmann
gas. Let us consider $h$ different particle species, with
masses $m_{i}$ and charges $q_{i}$, $i=1...h$. In this case, the
Boltzmann-Gibbs partition function is
\begin{equation}
Z\left(  \beta,\mu,V\right)  =\exp\left(  V\,\sum_{i=1}^{h}\Phi_{i}%
(\beta)\,\exp(\beta\mu q_{i})\right)  \;, \label{Boltz-gas}%
\end{equation}
where%
\begin{equation}
\Phi_{i}(\beta)=\frac{g_{i}}{2\pi^{2}}\frac{m_{i}^{2}}{\beta}K_{2}\left(
\beta m_{i}\right)  \;.
\end{equation}
Above, $g_{i}$ is the statistical factor of particle $i$ and $K_{2}\left(
z\right)  $ is a modified Bessel function.

The $N$-particle partition function is%
\begin{equation}
Z^{(N)}(\beta,\mu,V)=\frac{1}{N!}\left(  V\,\sum_{i=1}^{h}\Phi_{i}%
(\beta)\,\exp(\beta\mu q_{i})\right)  ^{N} \label{Boltz-Ngas}%
\end{equation}
and, from Eq.~(\ref{Pn}), the multiplicity distribution of the ideal Boltzmann
gas is seen to be Poissonian. The average number of particles is%
\begin{equation}
\overline{N}=n(\beta,\mu)V\;,
\end{equation}
where
\begin{equation}
n(\beta,\mu)=\,\sum\limits_{i=1}^{h}\Phi_{i}(\beta)\,\exp(\beta\mu q_{i})
\label{n}%
\end{equation}
is the particle density.

We meet a problem when we try to apply Tsallis'\ statistics to the ideal gas:
for $q>1$ the generalized partition function $Z_{q}$ is infinite. The integral
representation of $Z_{q}$, Eq.~(\ref{part-2}), diverges if the ideal gas
partition function (\ref{Boltz-gas}) is substituted in the integrand. This
happens because for $\beta\rightarrow0$,
\begin{equation}
\Phi_{i}(\beta)\simeq\frac{g_{i}}{\pi^{2}\beta^{3}}\;,
\end{equation}
so that $Z(x\beta,\mu,V)$ has an essential singularity at $x=0$ which cannot
be removed by any power of $x$ in $G\left(  x\right)  $. Therefore, the
$q$-statistics of an ideal relativistic gas cannot be defined --- there is no
ideal Tsallis gas with $q>1$.

\section{Relativistic van der Waals gas
\label{sec-vdW}}

When restricted to $N$ particles, the generalized ideal gas partition
function $Z_{q}^{(N)}$ has a well defined integral representation in terms
of the corresponding Boltzmann-Gibbs function (Eq.~(\ref{Boltz-Ngas})),
provided
\begin{equation}
N<\frac{1}{3(q-1)}\;. \label{Nmax}%
\end{equation}
It is the production of particles above this limit that causes the divergence
in the partition function of the Tsallis ideal gas. High multiplicities can be
suppressed if a repulsive interaction exists among the produced particles. A
simple way to include such interactions is to introduce a ``van der Waals''
excluded volume simulating the effect of hard-core potentials among particles.
The van der Waals gas model has been frequently used to analyze particle
abundances in high-energy heavy ion collisions \cite{Gorenstein,BraunM}.

Let $v_{0}$ be the excluded volume associated to a particle. The corresponding
partition function for $N~$particles is obtained by replacing the volume of
the system $V$ by $V-Nv_{0}$,%
\begin{equation}
Z^{\left(  N\right)  }\left(  \beta,\mu,V\right)  \rightarrow Z^{\left(
N\right)  }\left(  \beta,\mu,V-Nv_{0}\right)  \,\Theta\left(  V-Nv_{0}\right)
.
\end{equation}
The Heaviside $\Theta$-function limits the number of particles inside the
volume $V$ to $N<V/v_{0}$. The partition function for the relativistic van der
Waals gas is then \cite{Gorenstein}
\begin{equation}
Z\left(  \beta,\mu,V\right)  =\sum_{N}\frac{1}{N!}n(\beta,\mu)^{N}\left(
V-Nv_{0}\right)  ^{N}\,\Theta\left(  V-Nv_{0}\right)
\end{equation}
where $n(\beta,\mu)$ is the ideal gas density given in Eq.~(\ref{n}). In the
large $V$ asymptotic limit, this can be written as
\begin{equation}
Z\left(  \beta,\mu,V\right)  =\exp\left\{  \frac{V}{v_{0}}W\left[
v_{0}n(\beta,\mu)\right]  \right\}  \;,
\end{equation}
where $W(x)$ is the Lambert function \cite{Lambert1,Lambert2},
defined by the equation
\begin{equation}
W(x) e^{W(x)} = x  \;.
\end{equation}
Similarly, the generating function of the excluded volume gas is found to be
\begin{equation}
F(t) = \frac{1}{Z(\beta,\mu,V)}\exp\left\{  \frac{V}{v_{0}}
       W\left[t\,v_{0} n(\beta,\mu)\right]  \right\}  \;. 
\label{F-vdW}
\end{equation}

The Lambert function $W~(z)$ has the asymptotic behavior, in the limit
$z\rightarrow\infty$,
\begin{equation}
W\left(  z\right)  \simeq\ln z+\ln\ln z+\cdots\;.
\end{equation}
The essential singularity at $\beta\rightarrow0$ of the ideal Boltzmann gas
partition function is thus reduced to a power one,
\begin{equation}
Z\left(  \beta,\mu,V\right)  \sim\beta^{-3V/v_{0}}\;. \label{Z-asymp}%
\end{equation}
Therefore, the integral in
\begin{equation}
Z_{q}\left(  \beta,\mu,V\right)  =\int_{0}^{\infty}dxG\left(  x\right)
\exp\left\{  \frac{V}{v_{0}}W\left[  \,v_{0}n(x\beta,\mu)\right]  \right\}
\;,
\end{equation}
converges as long as
\begin{equation}
q<1+\frac{v_{0}}{3V}\;, \label{qmax}%
\end{equation}
as expected from Eq.~(\ref{Nmax}), since the hard core suppresses the
production of particles above the maximum number $V/v_{0}$. In this range of
$q$ values, the generating function of the multiplicity distribution in the
Tsallis - van der Waals gas is given by
\begin{equation}
F(t)=\frac{1}{Z_{q}\left(  \beta,\mu,V\right)  }\int_{0}^{\infty}%
dxG(x)\exp\left\{  \frac{V}{v_{0}}W\left[  t\,v_{0}n(x\beta,\mu)\right]
\right\}  \;. \label{F-TvdW}%
\end{equation}

\section{Tsallis and van der Waals corrections to the ideal gas
\label{sec-correc}}

In order to examine in more detail the effect of Tsallis statistics on the
multiplicity distribution, let us consider the case in which $q-1$ and $v_{0}$
are both small (respecting the limit of Eq.~(\ref{qmax})). The Lambert
function $W(z)$ has the series expansion \cite{Lambert1}
\begin{equation}
W(z) = \sum_{n=1}^{\infty} \frac{(-n)^{n-1}}{n!} \, z^{n}\;,
\end{equation}
so that for $v_{0}\rightarrow0$ we have%
\begin{equation}
W (tv_{0}n) \simeq tv_{0}n - ( tv_{0}n )^{2} + \cdots
\end{equation}
and the generating function for the van der Waals-Tsallis relativistic gas can
be written as
\begin{equation}
F(t)\simeq\frac{1}{Z_{q}}\int_{0}^{\infty}dxG\left(  x\right)  \exp\left\{
tVn(x\beta,\mu)\left[  1-tv_{0}n(x\beta,\mu)\right]  \right\}  \;.
\end{equation}
For $q\rightarrow1$, this integral can be calculated with the help of
Eq.~(\ref{q1}). To first order in $q-1$ and $v_{0}$ we obtain
\begin{eqnarray}
F(t) & \simeq& \exp \{ (t-1) Vn [1 + (q-1) \lambda (Vn\lambda-1)-2v_{0}n ]  
\nonumber \\
&  &  \mbox{} +  (t-1)^{2} (V n)^{2}[(q-1) \lambda^{2}/2 -v_{0}/V]  \}
\end{eqnarray}
where
\begin{equation}
\lambda(\beta,\mu)=-\frac{\beta}{n}\frac{\partial n}{\partial\beta}\;.
\end{equation}
Comparison to Eq.~(\ref{NBasymp}) shows that this is the generating function
of a negative binomial distribution, with average number of particles%
\begin{equation}
\overline{N}=Vn\left[  1+(q-1)\lambda(Vn\lambda-1)-2v_{0}n\right]
\end{equation}
and a (large) $k$-parameter given by%
\begin{equation}
\frac{1}{k}=(q-1)\lambda^{2}-2\frac{v_{0}}{V}\;. \label{k}%
\end{equation}
Strictly speaking, a negative binomial distribution is obtained only if $k>0$,
or
\begin{equation}
q>1+\frac{2v_{0}}{\lambda^{2}V}\;. \label{qmin}%
\end{equation}
In most cases of interest for hadron production the value of $\lambda
(\beta,\mu)$ is large, so that Eqs.~(\ref{qmin}) and (\ref{qmax}) can both
hold. If $q$ is bellow the limit of Eq.~(\ref{qmin}),\ $k$ is negative, and
the multiplicity distribution will have a binomial-like form. From
Eq.~(\ref{k}) we see that the effect of Tsallis statistics is to make the
multiplicity distribution wider than the Poisson distribution with the same
average number of particles. The effect of the van der Waals excluded volume
is to reduce the width relative to Poisson.

\section{Charged particle multiplicity in hadronic collisions}

Pions are the most copiously produced particles in high-energy hadronic
collisions. They come not only from the thermally equilibrated gas, but also
from the decay of heavier hadrons formed in the gas. A precise description of
particle multiplicities has to take into account how much these unstable
hadrons, called resonances, participate in the production of the observed
particles. In the present analysis, we intend to understand the basic
influence of the $q$-statistics on the particle multiplicity distribution.
Therefore, to simplify the picture, we will study the behavior of the
lowest-lying non-strange mesons, $\pi$, $\eta$, $\rho$ and $\omega$ in thermal
equilibrium. The $\eta$, $\rho$ and $\omega$\ mesons are resonances which
decay into pions. The essential features of the baryon-free hadronic gas can
be described by the mixture of these mesons, and the major conclusions of the
present work will not be changed by this simplification.

\begin{figure}[p]
\begin{center}
\includegraphics{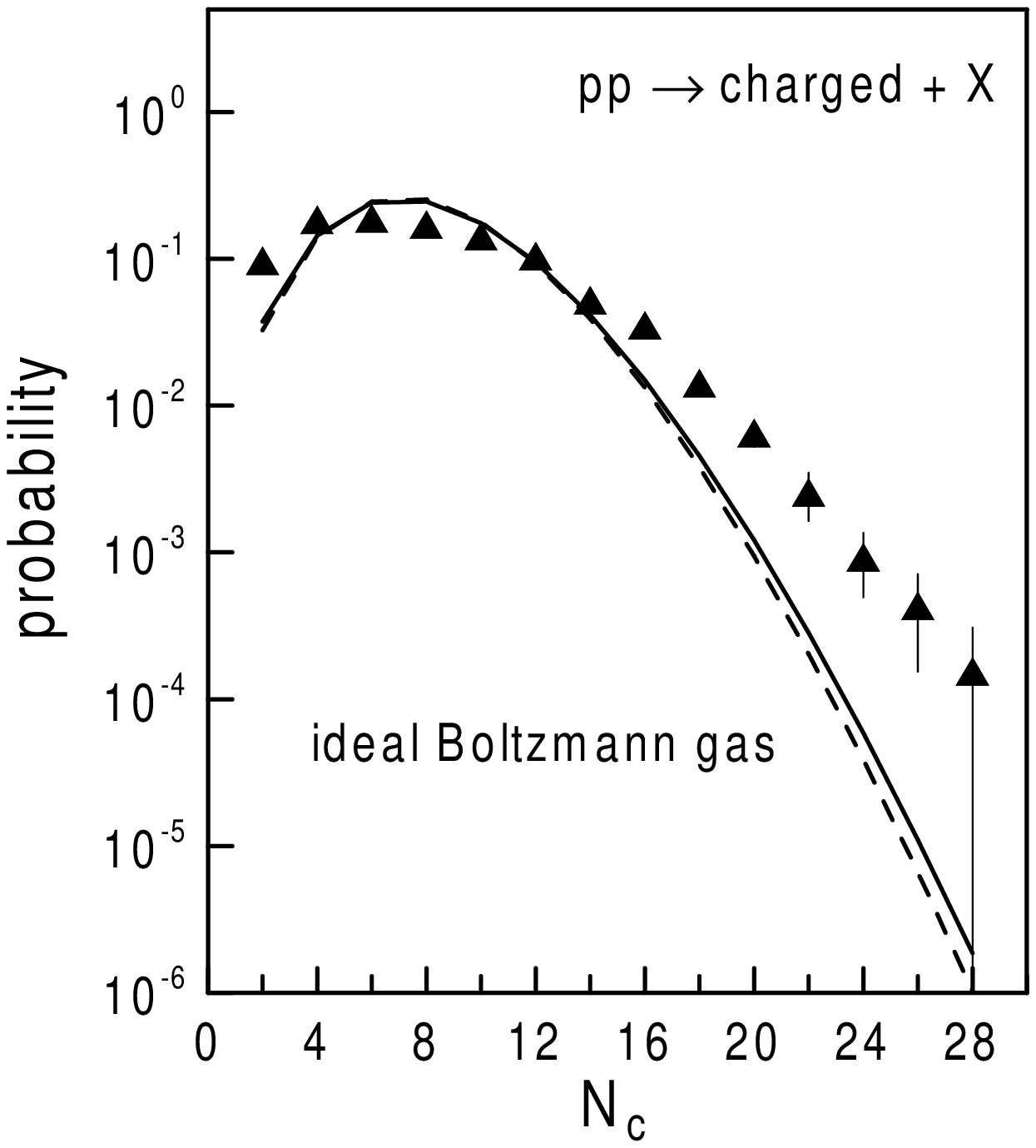}
\caption{Multiplicity distribution of charged particles produced in $pp$
collisions at 250~GeV/c. The lines are calculations based on the ideal
Boltzmann gas model. The solid line corresponds to $T=190$~MeV and
$V=18.84$~fm$^{3}$, the dashed line to $T=160$~MeV and 
$V=42.96$~fm$^{3}$ }
\label{fig-ideal}
\end{center}
\end{figure}

In Fig.~\ref{fig-ideal}, we show the multiplicity distribution of charged
particles produced in $pp$ collisions at momentum $p_{lab}=250$~GeV/c.
Triangles are data from the NA22 experiment \cite{NA22}. The average number of
charged particles is $\overline{N}=7.88\pm0.09$ and the width of the
distribution is $D=4.10 \pm 0.05$. 
The lines are multiplicity distributions calculated for the meson system
described above, treated as an ideal Boltzmann gas. 
Two different temperatures were considered --- $T=190$~MeV (solid line), 
and $T=160$~MeV (dashed line) --- corresponding to
the range of values found in analyses of $pp$ reactions \cite{BecattiniHeinz}. 
In both cases the volume of the gas was chosen such as to reproduce the
measured average multiplicity. For the higher temperature this gives
$V=18.8$~fm$^{3}$, and for the lower, $V=43.0$~fm$^{3}$. 
We assumed that the meson gas has a total electric charge
$Q=1$, which is the typical value in $pp$ collisions. Changing this charge to
$0$ or $2$ has only a small effect on the multiplicity distribution. 
All calculations were performed in the canonical framework.

Due to resonance decay, and exact charge conservation in the canonical 
ensemble, the calculated multiplicity does not follow exactly the Poisson 
distribution obtained in the grand canonical approach of Sec.~\ref{sec-ideal}. 
The deviation from Poisson is conveniently measured by the quantity 
\begin{equation}
\frac{1}{k}=\frac{D^{2}-\overline{N}}{\overline{N}^{2}}\;, 
\label{Poisson-dev}
\end{equation}
a definition motivated by the $k$-parameter of the negative binomial
distribution (compare with Eq.~(\ref{NB-k})). The ideal gas distributions
shown in Fig.~\ref{fig-ideal} have $k \sim 40$ and are, therefore, 
wider than Poisson. They are well approximated by a negative binomial
distribution with the same $k$ parameter. The large value of $k$ shows that
the deviation from Poisson caused by resonance decay and charge conservation
is relatively small. This is reflected in Fig.~\ref{fig-ideal}, where
multiplicity distributions constrained to have the same average present
similar widths, despite the very different gas temperatures and volumes. It
also indicates that the ideal gas model cannot explain the multiplicity data,
in particular its large width. For all reasonable values of $T$ and $V$
consistent with the experimental average multiplicity, the ideal gas shows
approximately the same narrow distribution. One may think that including more
hadronic resonances in the model improves the situation. 
However, Becattini \emph{et al}. \cite{Becattini2} find $k>10$ in their 
analysis of $e^{+}e^{-}$ data, which includes many resonances. 
Hadronic collisions produce somewhat wider multiplicity
distributions; the NA22 data shown in Fig.~\ref{fig-ideal} have $k=7$, and as
the energy increases $k$ becomes smaller \cite{Ward}. We thus conclude that
the multiplicity distribution of the ideal relativistic gas is not consistent 
with the hadron-hadron experimental data.

\begin{figure}[p]
\begin{center}
\includegraphics{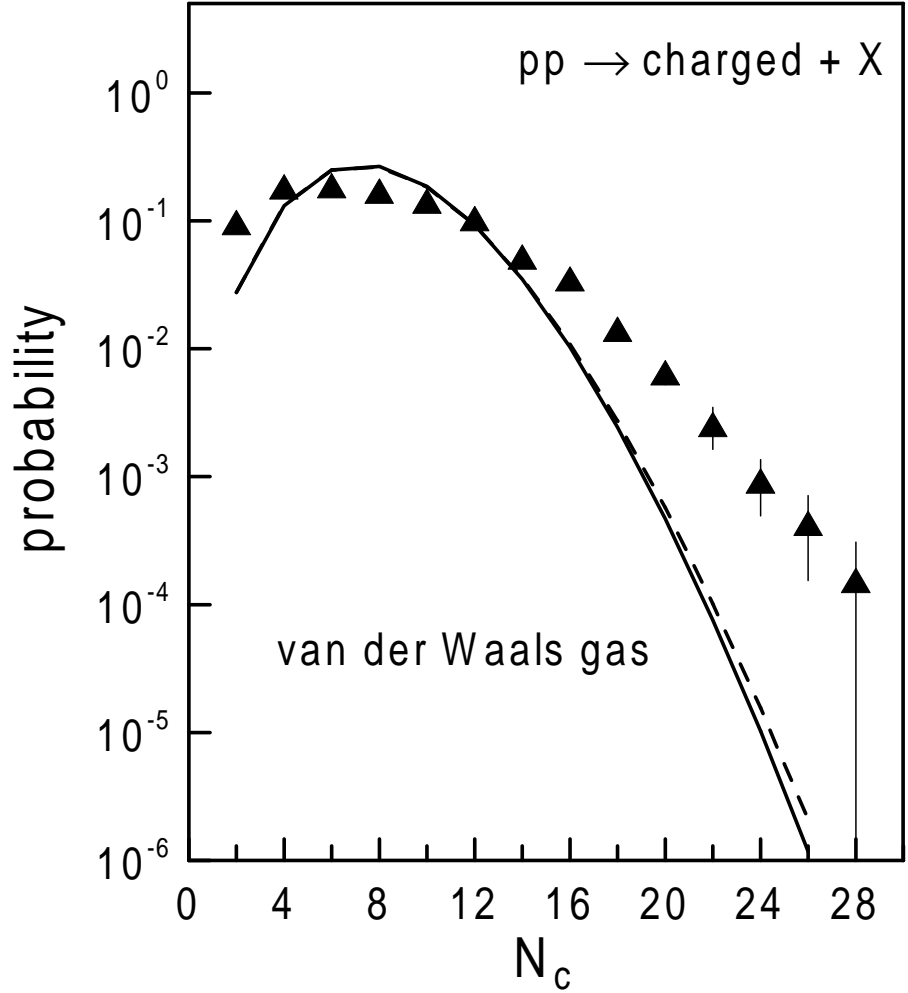}
\caption{Charged particle multiplicity distribution in $pp$ collisions at
250~GeV/c. The lines are multiplicity distribution of van der Waals gases with
$T=190$~MeV, $V=24.46$~fm$^{3}$ (solid), and $T=160$~MeV, $V=49.21$~fm$^{3}$
(dashed). In both calculations the excluded volume is $v_{0}=0.368$~fm$^{3}$.}%
\label{fig-vdW}%
\end{center}
\end{figure}

The situation doesn't get better if the meson system is treated as a 
relativistic van der Waals gas. 
This is seen in Fig.~\ref{fig-vdW}, where the NA22 data is compared to the 
canonical multiplicity distribution of the meson gas with excluded 
volume $v_{0}=0.368$~fm$^{3}$. 
Again two temperatures, $T=190$~MeV (solid line) and
$160$~MeV (dashed line), were considered, with volumes $V=24.5$~fm$^{3}$ and
$49.2$~fm$^{3}$, respectively, determined by the experimental average
multiplicity. The excluded volume reduces the width of the multiplicity
distribution, relative to the ideal gas with the same average multiplicity; 
we have $k \sim 100$ for the van der Waals gas distributions shown in
Fig.~\ref{fig-vdW}. The width reduction occurs for any value of $v_{0}$.
Therefore, the van der Waals gas model cannot explain the experimental
multiplicity distribution; it is even less satisfactory than the ideal 
gas model.

\begin{figure}[p]
\begin{center}
\includegraphics{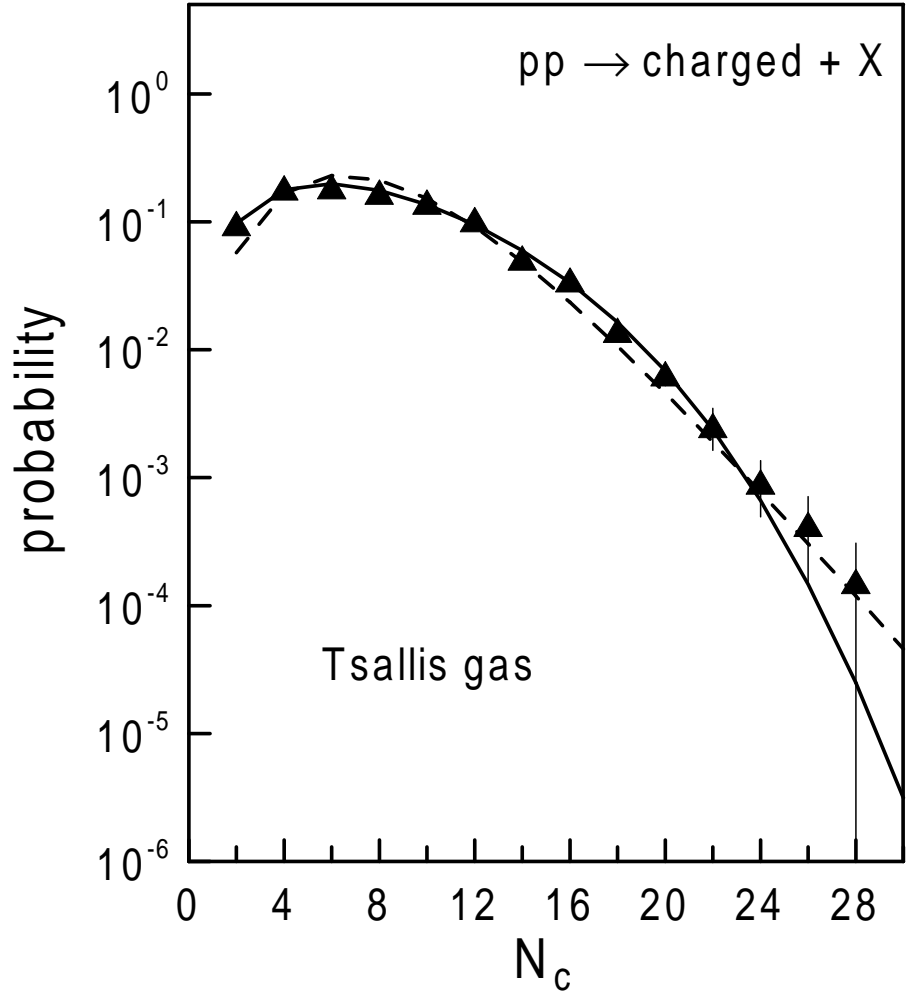}
\caption{Charged particle multiplicity distribution in $pp$ collisions at
250~GeV/c. The lines are multiplicity distribution of Tsallis gases with
$\beta^{-1}=190$~MeV, $V=9.13$~fm$^{3}$, $q=1.0097$ (solid), and $\beta
^{-1}=160$~MeV, $V=31.0$~fm$^{3}$, $q=1.0030$ (dashed). In both cases
$v_{0}=0.368$~fm$^{3}$.}%
\label{fig-Tsallis}%
\end{center}
\end{figure}

Tsallis statistics for $q>1$ produces multiplicity distributions that 
are wider than the Boltzmann-Gibbs ones (see Sec.~\ref{sec-correc}). 
In Fig.~\ref{fig-Tsallis}, we show some examples of the canonical multiplicity
distribution obtained in $q$-statistics. The solid line corresponds to a
gas with $\beta^{-1}=190$~MeV, $V=9.13$~fm$^{3}$, $q=1.0097$, and the dashed
line to $\beta^{-1}=160$~MeV, $V=31.0$~fm$^{3}$, $q=1.0030$. In both
calculations the excluded volume is $v_{0}=0.368$~fm$^{3}$. We note that a even
a small deviation of $q$ from unity causes a substantial broadening of the
distribution, taking it much closer to the experimental data.
Thus, it seems possible to give a thermostatistical description of 
high-energy hadronic multiplicities, provided nonextensivity effects
are taken into account.

\section{Concluding remarks}

In the present work we have studied the effect of the Tsallis nonextensive
statistics (with $q>1$) on the multiplicity distribution of particles in a
high-energy gas. In order to avoid divergences in the generalized partition
function of the ideal gas, a van der Waals hard-core interaction was
introduced as a physical regularization procedure. Nonextensive statistics
gives rise to multiplicity distributions that are broader than the
Boltzmann-Gibbs ones. In the small $q-1$ limit, the Tsallis
multiplicity distribution can be approximated by a negative binomial
distribution, with $k$-parameter proportional to $(q-1)^{-1}$. We 
developed a simple meson gas model in order to compare the different
statistics to experimental data from hadronic collisions. The multiplicity
distribution obtained with the Boltzmann-Gibbs statistics has a too narrow
width compared to the experimental results. A small increase of $q$ from unity
makes the distribution considerably wider, and a good agreement
with the data can be obtained within Tsallis statistics.
These results provide an interesting perspective on why hadronic multiplicities
have a negative binomial distribution, and suggest that nonextensive thermal
phenomena play an important role in high-energy collisions.

In order to get a more complete picture of the nonextensive thermostatistical 
effects in hadron production, we need to extend our analysis using a
larger resonance gas, and including other observables such as particle ratios,
transverse momentum spectra, and correlations. 
Work along these lines is in progress.

\medskip
We wish to thank C. Tsallis and G. Wilk for fruitful discussions.
This work was partially supported by FAPERJ and CNPq.

\end{document}